\documentclass[10pt]{article}
\usepackage{graphicx,amsmath,amssymb,url,verbatim,SIGACTNews}
\usepackage[tight]{subfigure}
\usepackage{verbatim,url}
\usepackage{graphicx, color}
\usepackage{color}
\usepackage{colortbl}
\usepackage[table]{xcolor}

\newcommand\colortext[1]{{{#1}}}
\newcommand\bluetext[1]{{{#1}}}
\newcommand\revised[1]{{{#1}}}

\begin{document}

\title{An Overview of Codes Tailor-made for \revised{Better Repairability in} Networked Distributed Storage Systems}
\author{Anwitaman Datta, Fr\'ed\'erique Oggier}
\renewcommand{\baselinestretch}{1}
\maketitle
\thispagestyle{empty}

\begin{abstract}
The increasing amount of digital data generated by today's society asks for better storage solutions. This survey looks at a new generation of coding techniques designed specifically for the maintenance needs of networked distributed storage systems \bluetext{(NDSS)}, trying to reach the best compromise among storage space efficiency, fault-tolerance, and maintenance overheads. Four families of codes, namely, pyramid, hierarchical, regenerating and \revised{locally repairable codes such as} self-repairing codes, \revised{along with a heuristic of cross-object coding to improve repairability in NDSS} are presented at a high level. \revised{The code descriptions are accompanied with simple examples} emphasizing the main ideas behind each of \revised{these code families}. We discuss their pros and cons before concluding with a \revised{brief and preliminary} comparison. This survey deliberately excludes \bluetext{technical details and does not contain an exhaustive list of code constructions}. Instead, it provides an overview of the major novel code families in a manner easily accessible to a broad audience, by presenting the big picture of advances in coding techniques for \bluetext{maintenance of NDSS}.
\\
\textbf{Keywords: coding techniques, \bluetext{networked distributed} storage systems, hierarchical codes, pyramid codes, regenerating codes, \revised{locally repairable codes,} self-repairing codes, \revised{cross-object coding}.}
\end{abstract}


\section{Introduction}
\colortext{We live in an age of data deluge}. A study sponsored by the information storage company \emph{EMC} estimated
that the world's data is more than doubling every two years, reaching 1.8 zettabytes (1 ZB = $10^{21}$ Bytes) of data to be stored in
2011.\footnote{\url{http://www.emc.com/about/news/press/2011/20110628-01.htm}}
This includes various digital data continuously being generated by
individuals as well as business and government organizations, who all need
scalable solutions to store data reliably and securely.

\colortext{Storage technology has been evolving fast in the last quarter of a century to meet the numerous challenges posed in storing an increasing amount of data and catering to diverse applications with different workload characteristics. In 1988, RAID (Redundant Arrays of Inexpensive Disks) was proposed \cite{raid}, which combines multiple storage disks (typically from two to seven) to realize a single logical storage unit. Data is stored redundantly, using replication, parity, or more recently erasure codes. Such redundancy makes a RAID logical unit significantly more reliable than the individual constituent disks. Besides meeting cost effective reliable storage, RAID systems provide good throughput by leveraging parallel I/O at the different disks, \bluetext{and, more recently,} geographic distributions of the constituent disks to achieve resilience against local events (such as a fire) that could cause correlated failures.}

\colortext{While RAID has evolved and stayed an integral part of storage solutions to date, new classes of storage technology have emerged, where multiple logical storage units (simply referred to as `storage nodes') are assembled together to scale out the storage capacity of a system. The massive volume of data involved means that it would be extremely expensive, if not impossible, to build single pieces of hardware
with enough storage as well as I/O capabilities.
By the term `networked', we refer to these storage systems that pool resources from multiple interconnected storage nodes, which in turn may or not use RAID. The data is distributed across these \bluetext{interconnected} storage units and hence \bluetext{the name} `networked distributed storage systems' (NDSS). \revised{It is worth emphasizing at this juncture that though the term `RAID' is now also used in the literature for NDSS environments, for instance, HDFS-RAID \cite{hdfs} and `distributed RAID'\footnote{http://www.disi.unige.it/project/draid/distributedraid.html}, but in this article we use the term RAID to signify traditional RAID systems where the storage nodes are collocated, and the number of parity blocks per data object is few, say one or two (RAID-1 to RAID-6)}. Unlike in traditional RAID systems where the storage disks are collocated, all data objects are stored in the same set of storage disks, and these disks share an exclusive communication bus within a stand-alone unit, in NDSS, a shared interconnect is used across the storage nodes, and different objects may be stored across arbitrarily different (possibly intersecting) subsets of storage nodes, and thus there is competition and interference in the usage of the network resources.}

\colortext{
NDSS come in many flavors such as data centers and peer-to-peer (P2P) storage/backup systems. While data centers comprise thousands of compute and storage nodes, individual clusters such as that of Google File System (GFS) \cite{GFS} are formed out of hundreds up to thousands of nodes. P2P systems like Wuala,\footnote{\revised{The current deployment of Wuala (\url{www.wuala.com}) no longer uses a hybrid peer-to-peer architecture.}} in contrast, formed swarms of tens to hundreds of nodes for individual files or directories, but would distribute such swarms arbitrarily out of hundreds of thousands of peers.}

\colortext{
While P2P systems are geographically distributed and connected through an arbitrary topology, data center interconnects have well defined topologies and are either collocated or distributed across a few geographic regions. Furthermore, individual P2P nodes may frequently go offline and come back online (temporary churn), creating unreliable and heterogeneous connectivity. On the contrary, data centers use dedicated resources with relatively infrequent temporary outages.}

\colortext{
Despite these differences, NDSS share several common characteristics. While I/O of individual nodes continues to be a potential bottleneck, available bandwidth, both at the network's edges and within the interconnect becomes a critical shared resource. Also, given the system scale, failure of a significant subset of the constituent nodes, as well as other network components, is the norm rather than the exception. To enable a highly available overall service, it is thus essential to tolerate both short-term outages of some nodes and to provide resilience against permanent failures of individual components. Fault-tolerance is achieved using redundancy, while long-term resilience relies on replenishment of lost redundancy over time.}

A common practice to realize redundancy is to keep three copies of \colortext{an object to be stored} (called 3-way replication): when one copy is lost, the second copy is used to regenerate
the first one, and hopefully, not both the \bluetext{remaining} copies are lost before the repair
is completed. There is of course a price to pay: redundancy naturally reduces the
efficiency, or alternatively put, increases the overheads of the storage infrastructure.
The cost for such an infrastructure should be estimated not only in terms of the hardware, but also of real estate and maintenance of a data center. A US Environmental Protection Agency report of
2007\footnote{http://arstechnica.com/old/content/2007/08/epa-power-usage-in-data-centers-could-double-by-2011.ars} indicates that the US used 61 billion kilowatt-hours of power for data centers and
servers in 2006. That is 1.5 percent of the US electricity use, and it cost the companies that paid those bills more than \$4.5 billion.

There are different ways to reduce these expenses, starting from the physical media, which has
witnessed a continuous shrinking of physical space and cost per unit of data, as well as reductions in terms of cooling needs.
This article focuses on a different aspect, that of the trade-off between fault-tolerance and efficiency in storage space utilization via coding techniques, or more precisely erasure codes.

\begin{figure}[htbp]
  \begin{center}
    \includegraphics[scale=0.75]{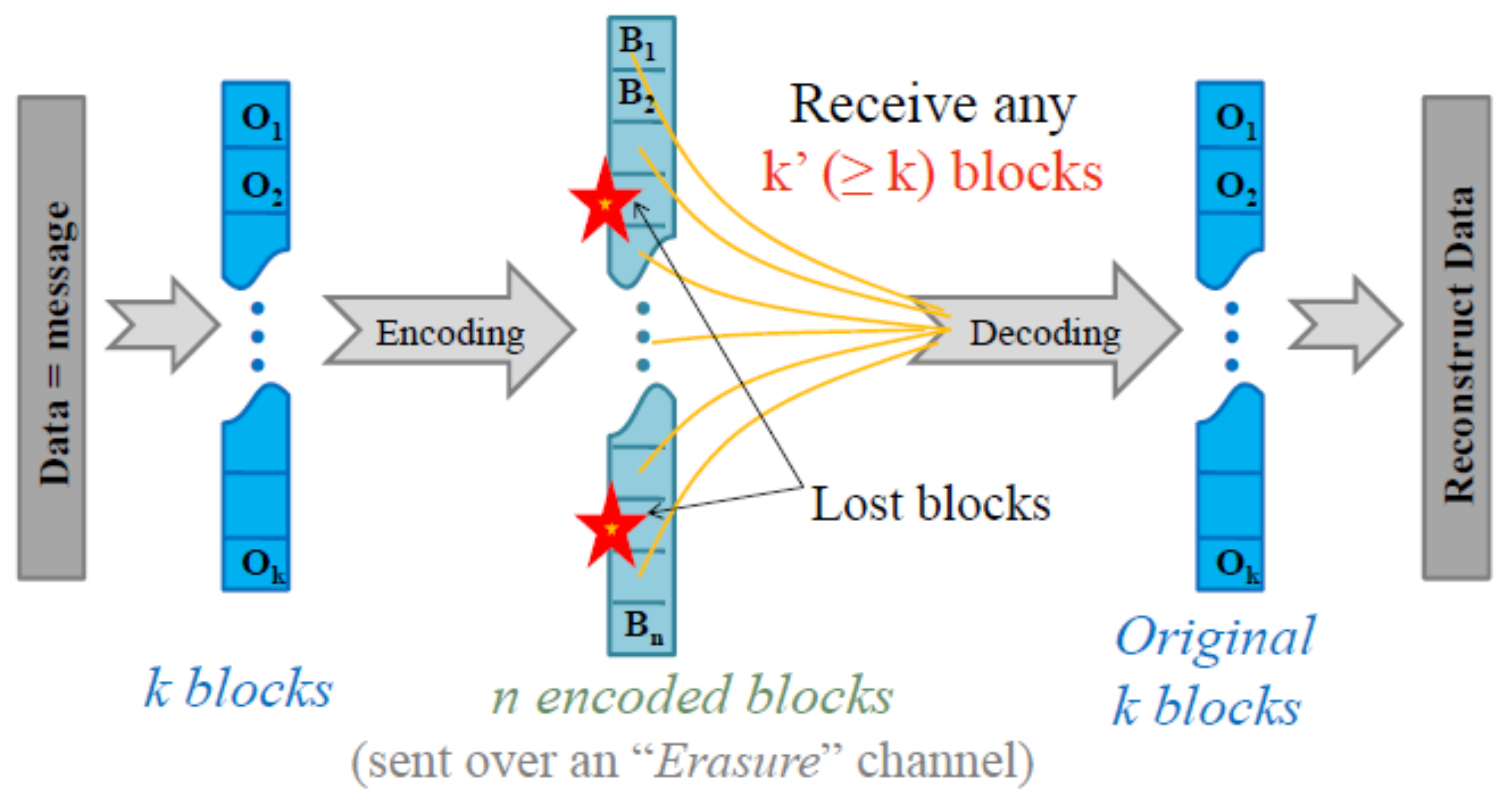}
    \caption{Coding for erasure channels: a message of $k$ symbols is encoded into $n$ fragments before transmission over an erasure channel. As long as at least $k'\geq k$ symbols arrive at destination, the receiver can decode the message.}
  \label{fig:ecforcomm}
  \end{center}
\end{figure}

\colortext{
An erasure code $EC(n,k)$ transforms a sequence of $k$ symbols into a longer sequence of $n>k$ symbols. Adding extra $n-k$ symbols helps in recovering the original data in case some of the $n$ symbols are lost. An $EC(n,k)$ induces a \bluetext{$n/k$ overhead}.
Erasure codes were designed for data transmitted} over a noisy channel, where coding is used to append redundancy to the transmitted signal to help the receiver recover the intended message, even when some symbols are erasured/corrupted by noise (see Figure~\ref{fig:ecforcomm}).
\colortext{
Codes offering the best trade-off between redundancy and fault-tolerance, called maximum distance separable (MDS) codes, tolerate $n-k$ erasures, that is, no matter which group of $n-k$ symbols are lost, the original data can be recovered.
The simplest examples are the repetition code $EC(n,1)$ (given $k=1$ symbol, repeat it $n$ times), which is the same as replication, and the parity check code $EC(k+1,k)$ (compute one extra symbol which is the sum of the first $k$ symbols).
The celebrated Reed-Solomon codes \cite{ReedSolomon} are another instance of such codes: consider a sequence of $k$ symbols as a degree $k-1$ polynomial, which is evaluated in $n$ symbols. Conversely, given $n$ symbols, or in fact at least (any) $k$ symbols, it is possible to interpolate them to recover the polynomial \bluetext{and} decode the data. Think of a line in the plane. Given any $k=2$ or more points, the line is completely determined, while with only one point, the line is lost.}

\colortext{This same storage overhead/fault tolerance trade-off has also long been studied in the context of RAID storage units. While RAID 1 uses replication, subsequent RAID systems integrate parity bits, and Reed-Solomon codes can be found in RAID 6. Notable \revised{examples of new codes designed to suit the peculiarities of RAID systems include weaver codes \cite{weaver}, array codes \cite{array} as well as other heuristics \cite{xorcodes}. Optimizing the codes for the nuances of RAID systems, such as physical proximity of storage devices leading to clustered failures are natural aspects \cite{yuval} gaining traction.}}
\revised{Note that even though we do not detail here those codes optimized for traditional RAID systems, they may nonetheless provide some benefits in the context of NDSS, and vice-versa.}

\colortext{A similar evolution has been observed in the world of NDSS, and a wide-spectrum of NDSS have started to adopt erasure codes: for example, the new version of Google's file system, Microsoft's Windows Azure Storage \cite{Calder-Azure} as well as other storage solution companies such as CleverSafe\footnote{\url{http://www.cleversafe.com/}} and Wuala. This has happened due to a combination of several factors, including years of implementation refinements, ubiquity of significantly powerful but cheap hardware, as well as the sheer scale of the data to be stored.}

\colortext{
We will next elaborate how erasure codes are used in NDSS, and while MDS codes are optimal in terms of fault-tolerance and storage overhead tradeoffs, why there is a renewed interest in the coding theory community to design new codes that take into account \bluetext{maintenance} of NDSS explicitly.}


\section{Networked Distributed Storage Systems}

\colortext{In an NDSS, if one object is stored using an erasure code and each `encoded symbol' is stored at a different node, then the object stays available as long as the number of node failures does not exceed the code recovery capability.}

\begin{figure*}[htbp]
    \begin{center}
    \subfigure[\label{fig:ecstorage}Data retrieval: as long as $k'\geq k$ nodes are alive, the object can be retrieved.]{\includegraphics[scale=0.5]{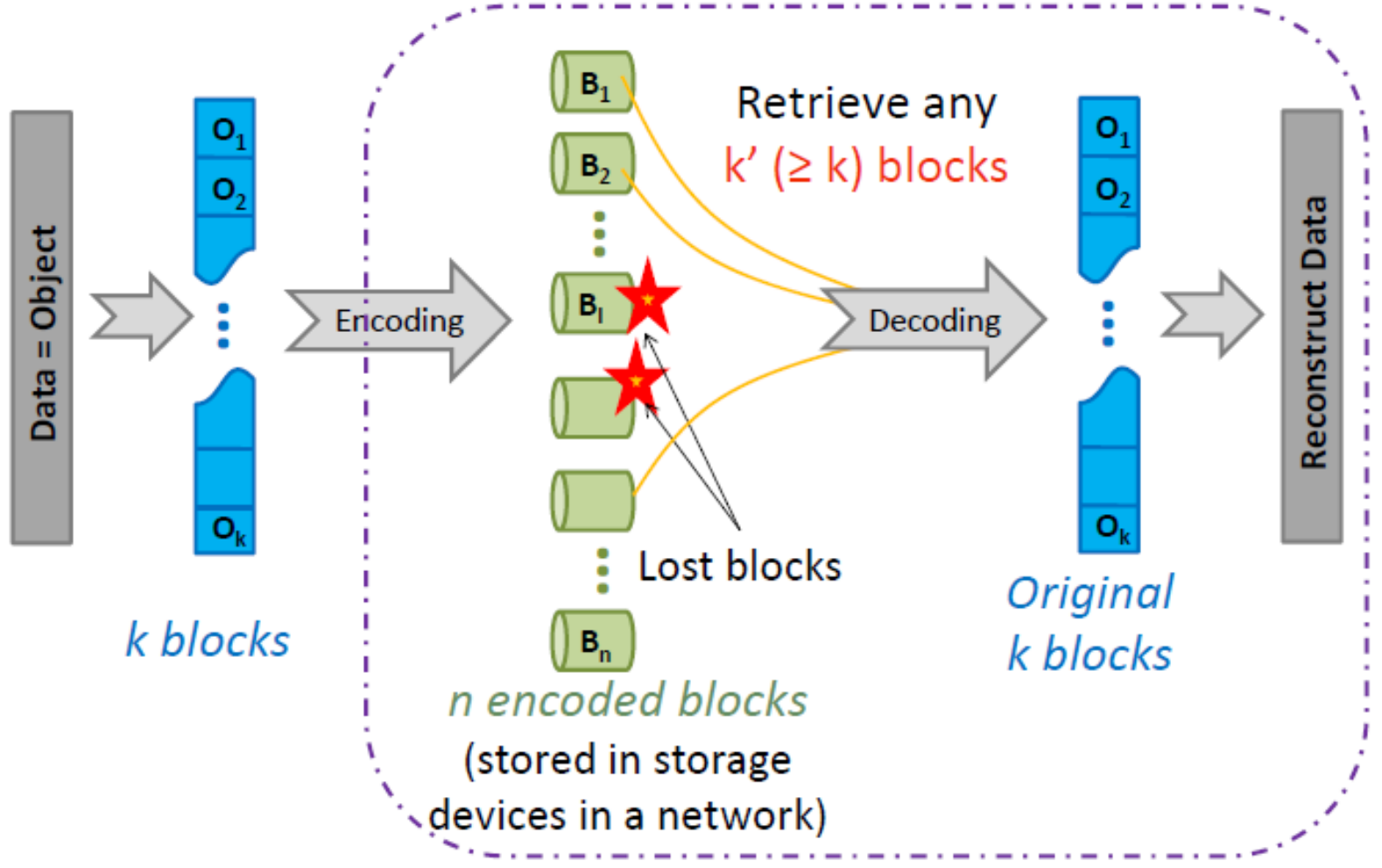}} \hspace{2mm}
    \subfigure[\label{fig:ecrepair}Node repair: one node has to recover the object, re-encode it, and then distribute the lost blocks to the new nodes.]{\includegraphics[scale=0.5]{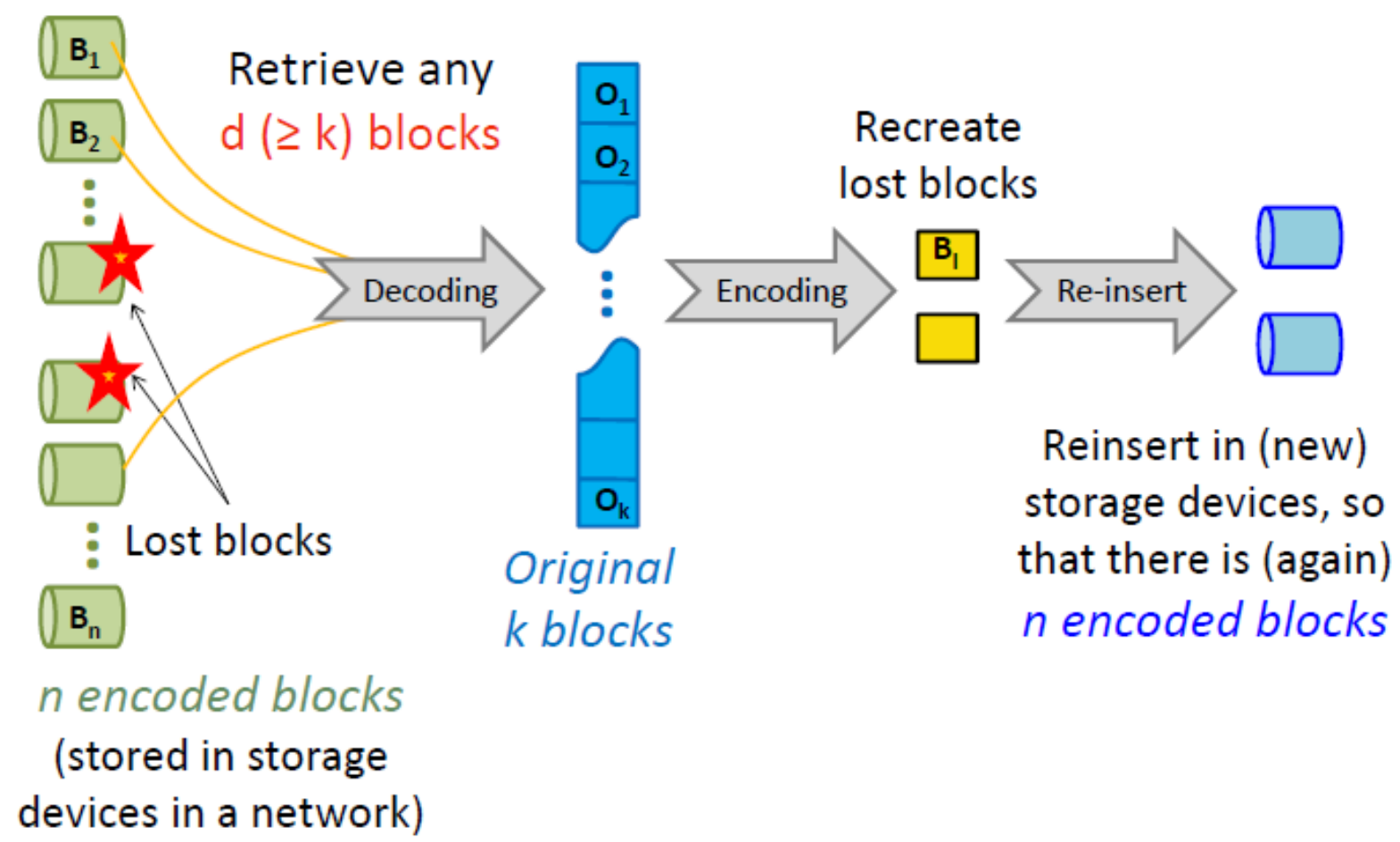}}
  \end{center}
  \caption{Erasure coding for NDSS: the object to be stored is cut into $k$, then encoded into
  $n$ fragments, given to different storage nodes. Reconstruction of the data is shown on the left, while repair after node failures is illustrated on the right.}
  \label{fig:ECdrawback}
\end{figure*}

\colortext{
Now let individual storage nodes fail \revised{according to an i.i.d. random process with the failure probability being $f$}. The expected number of independent node failures is binomially distributed, hence the probability of losing an object with an $EC(n,k)$ MDS erasure code is $\sum_{j=1}^k {n \choose n-k+j} f^{n-k+j}(1-f)^{k-j}$. In contrast, it is $f^r$ with $r$-way replication. For example, if the probability of failure of individual nodes is $f=0.1$, then for the same storage overhead of 3, corresponding to $r=3$ for replication and to an $EC(9,3)$ erasure code, the probabilities of losing an object are $10^{-3}$ and $\sim 3\cdot 10^{-6}$ respectively.
\bluetext{Such resilience analysis} illustrates the high fault-tolerance that erasure codes provide. \revised{Using erasure codes, however, means that a larger number of storage nodes are involved in storing individual data objects}.}

\colortext{There is however a fundamental difference between a communication channel, where erasures occur once during
transmission, and an NDSS, where faults accumulate over time, threatening data availability in the long run.}

\colortext{Traditionally, erasure codes were not designed to reconstruct subsets of arbitrary encoded blocks efficiently. When a data block encoded by an MDS erasure code is lost and has to be recreated, one would typically first need data equivalent in amount to recreate the whole object in one place (either by storing a full copy
of the data, or else by downloading an adequate number of encoded blocks), even in order to recreate a single encoded block, as illustrated in Figure~\ref{fig:ECdrawback}.}

\colortext{In recent years, the coding theory community has thus focused on designing codes which better suit NDSS nuances, particularly with respect to replenishing lost redundancy efficiently. \revised{The focus of such works has been on (i) bandwidth, which is typically a scarce resource in NDSS, (ii) the number of storage nodes involved in a repair process, (iii) the number of disk accesses (I/O) at the nodes facilitating a repair, and (iv) the repair time, since delay in the repair process may leave the system vulnerable to further faults. Note that these aspects are often interrelated.}
}
\colortext{
There are numerous other aspects, such as data placement,  meta-information management to coordinate the network, as well as interferences among multiple objects contending for resources, to name a few prominent ones, which all together determine an actual system's performance. The novel codes we describe next are yet to go through a comprehensive benchmarking across this wide spectrum of metrics. Instead, we hope to make \revised{these early and mostly theoretical} results accessible to practitioners, in order to accelerate the process of such further investigations.}

\colortext{ Thus the rest of this article assumes a
network of $N$ nodes, storing one object of size $k$, encoded into
$n$ symbols, also referred to as encoded blocks or fragments, each
of them being stored at distinct $n$ nodes out of the $N$ choices.
When a node storing no symbol \revised{corresponding to the object being repaired} participates in the repair process by downloading data from nodes owning data (also called {\em live
nodes}), it is termed a {\em newcomer}. Typical values of $n$ and
$k$ depend on the environments considered: for data centers, the
number of temporary failures is relatively low, thus small $(n,k)$
values such as $(9,6)$ or \bluetext{$(13,10)$} (with respective overheads of
1.5 and 1.3) are generally fine \bluetext{\cite{hdfs}}. In
P2P systems such as Wuala, larger parameters \bluetext{like $(517,100)$
are desirable to
guarantee availability since
nodes frequently go temporarily offline.}} \colortext{When discussing the repair properties
of a code, it is also important to distinguish which repair
strategy is best suited: in P2P systems, \bluetext{a lazy approach (where several
failures are tolerated before triggering repair) can avoid unnecessary repairs since nodes may be temporarily offline. Data centers might instead opt for immediate repairs.} Yet, proactive repairs can lead to cascading
failures\footnote{For example
\url{http://storagemojo.com/2011/04/29/amazons-ebs-outage/}}. Thus
in all cases, ability to repair multiple faults simultaneously is
essential.}

\colortext{
In summary, codes designed to optimize the maintenance process should take into account different code parameters, repair strategies, the ability to replenish single as well as multiple lost fragments, and repair time. Recent coding works aimed in particular at:}

\textbf{(i)} Minimize the absolute amount of data transfer needed to recreate one lost encoded block at a time when storage nodes fail. \emph{Regenerating codes} \cite{DGWK-journal} form a new family of codes achieving the minimum possible repair bandwidth (per repair) given an amount of storage per node, where the optimal storage-bandwidth trade-off is determined using a network coding inspired analysis, assuming that each \bluetext{new-coming} node contacts $d \geq k$ arbitrary live nodes for each repair. Regenerating codes, like MDS erasure codes, allow data retrievability from any arbitrary set of $k$ nodes. Collaborative regenerating codes \cite{Shum-ICC, KLS} are a generalization allowing \bluetext{simultaneous repair of} multiple faults.

\textbf{(ii)} Minimize the number of nodes to be contacted for recreating one encoded block, \revised{referred to as fan-in}. \colortext{Reduction in the number of nodes needed for one repair typically increases the number of ways repair may be carried out, thus avoiding bottlenecks caused by stragglers. It also makes multiple parallel repairs possible, all in turn translating into faster system recovery.} \revised{To the best of our knowledge, \emph{self-repairing codes} \cite{OD-infocom} were the first instances of $EC(n,k)$ code families achieving a repair fan-in of 2 for up to $\frac{n-1}{2}$ simultaneous and arbitrary failures. Since then, such codes have become a popular topic of study under the nomenclature of `locally repairable codes' - the name being reminiscent of a relatively well established theoretical computer science topic of locally decodable codes. Other specific instances of locally repairable code families such as \cite{OD-itw,gopalan,ankit}, as well as study of the fundamental trade-offs and achievability of such codes \cite{henk} have commenced in the last years.}

\revised{Local repairability come at a price, since either nodes store the minimum possible amount of data, in which case the MDS property has to be sacrificed \colortext{(if one encoded symbol can be repaired from other two, any set of $k$ nodes including these 3 nodes will not \bluetext{be adequate} to reconstruct the data)}, or the amount of data stored in each node has to be increased.} A \bluetext{\emph{resilience analysis}} of self-repairing codes \cite{OD-infocom} has shown that object retrieval is little impaired by \bluetext{it,} 
and in fact, the MDS property might not be as critical for NDSS as it is for communication, since NDSS have the option of repairing data.

There are \bluetext{other} codes which fall somewhere `in between' these extremes. Prominent among these are hierarchical and pyramid codes which we summarize first before taking a closer look at regenerating and locally repairable codes.


\section{Hierarchical and Pyramid codes}
\begin{figure}[htbp]
  \begin{center}
\includegraphics[scale=0.75]{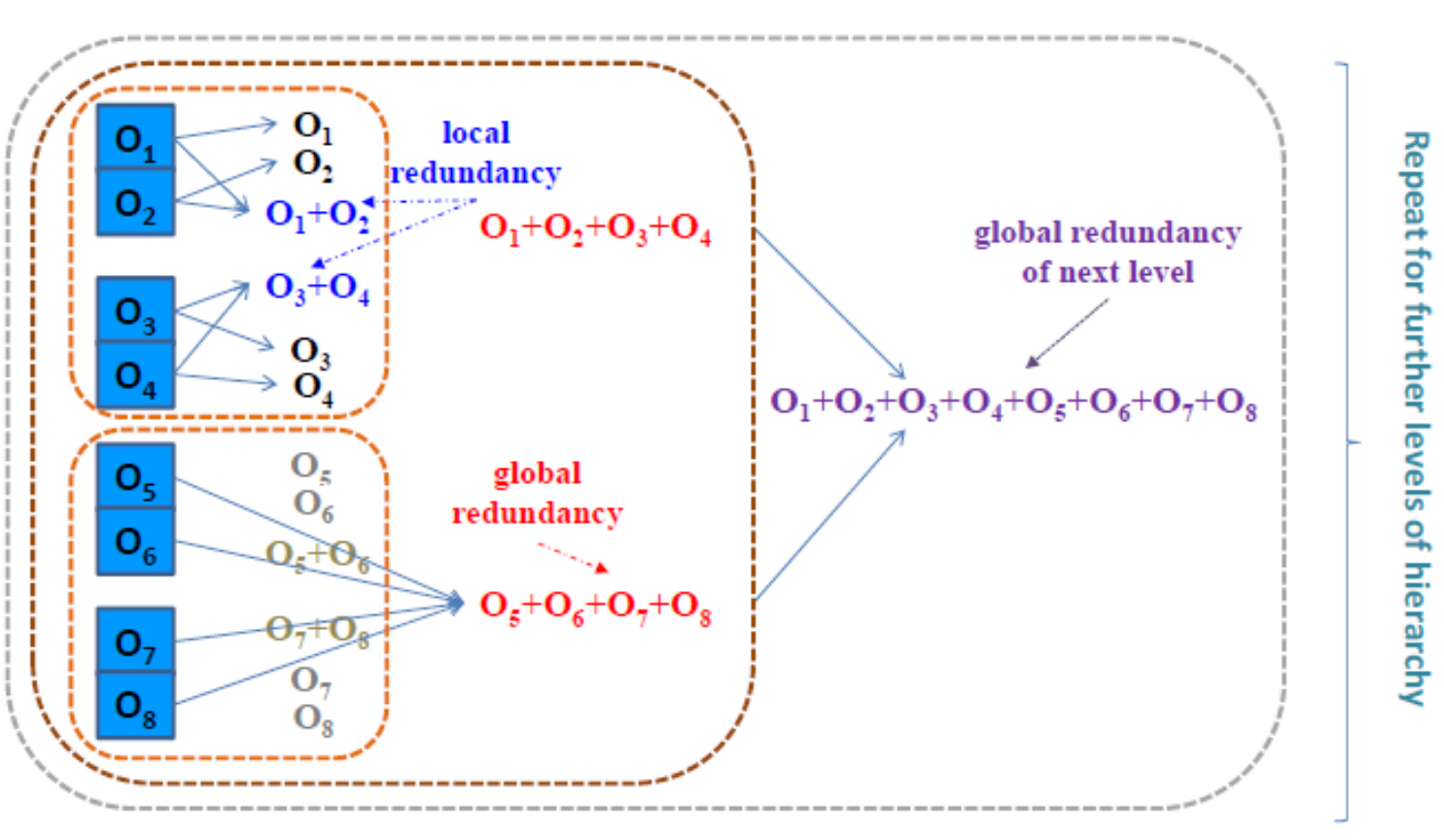}
    \caption{Hierarchical codes.}
  \label{fig:hierarchical}
 \end{center}
\end{figure}

\colortext{Consider an object comprising eight data blocks ${\bf o}_1,\ldots,{\bf o}_8$. Create three encoded fragments ${\bf o}_1$, ${\bf o}_2$ and ${\bf o}_1+{\bf o}_2$ using the first two blocks, and repeat the same process for blocks ${\bf o}_{2j+1}$ and ${\bf o}_{2j+2}$ (for $j=1...3$). One can then build another layer of encoded blocks, namely ${\bf o}_1+{\bf o}_2+{\bf o}_3+{\bf o}_4$ and ${\bf o}_5+{\bf o}_6+{\bf o}_7+{\bf o}_8$. The fragment ${\bf o}_1+{\bf o}_2$ may be viewed as providing \emph{local redundancy}, while ${\bf o}_1+{\bf o}_2+{\bf o}_3+{\bf o}_4$ achieves \emph{global redundancy}. The same idea can be iterated to build a hierarchy (Figure \ref{fig:hierarchical}), where the next level global redundant fragment is ${\bf o}_1+{\bf o}_2+{\bf o}_3+{\bf o}_4+{\bf o}_5+{\bf o}_6+{\bf o}_7+{\bf o}_8$.}

\colortext{Consequently, when some of the encoded fragments are lost, localized repair is attempted, and global redundancy is used only if necessary. For instance, if the node storing ${\bf o}_1$ is lost, then nodes storing ${\bf o}_2$ and ${\bf o}_1+{\bf o}_2$ are adequate for repair. However, if nodes storing ${\bf o}_1$ and ${\bf o}_1+{\bf o}_2$ are both lost, one may first reconstruct ${\bf o}_1+{\bf o}_2$ by retrieving ${\bf o}_1+{\bf o}_2+{\bf o}_3+{\bf o}_4$ and ${\bf o}_3+{\bf o}_4$, and then rebuild ${\bf o}_1$.}

\colortext{This basic idea can be extended to realize more complex schemes, where (any standard) erasure coding technique is used \revised{in a \emph{bottom-up} manner} to create local and global redundancy at a level, and the process is iterated. That is the essential idea behind \emph{Hierarchical codes} \cite{hierarchical}.} \revised{For the same example, one may also note that if both ${\bf o}_1$ and ${\bf o}_2$ are lost, then repair is no longer possible. This illustrates that the different encoded pieces have unequal importance. Because of such assymmetry, the resilience of such codes have only been studied with simulations in \cite{hierarchical}.}

\revised{In contrast, \emph{Pyramid codes} \cite{pyramid} were designed in a top-down manner, but aiming again to have local and global redundancy to provide better fault-tolerance and improve read performance by trading storage space efficiency for access efficiency. Such local redundancy can naturally be harnessed for efficient repairs as well. A new version of Pyramid codes, where the coefficients used in the encoding have been numerically optimized, namely Locally Reconstructable Codes \cite{AzureLRC} has more recently been proposed and is being used in the Azure \cite{Calder-Azure} system.}

\revised{We use an example to illustrate the design of a simple Pyramid code.
Take an $EC(11,8)$ MDS code, say a Reed-Solomon code with generator matrix $G$, of the form
\[
[x_1,\ldots,x_{11}]=[{\bf o}_1,\ldots,{\bf o}_8,c_1,c_{2},c_{3}].
\]}

\revised{A Pyramid code can be built from this base code, by retaining the pieces ${\bf o}_1,\ldots,{\bf o}_8$, and two of the other pieces (without loss of generality, lets say, $c_2,c_3$).}

\revised{Additionally, split the data blocks into two groups ${\bf o}_1,\ldots,{\bf o}_4$ and ${\bf o}_5,\ldots,{\bf o}_8$, and compute some more redundancy coefficients for each of the two groups, which is done by picking a first symbol $c_{1,1}$ corresponding to $c_1$ by setting ${\bf o}_5=\ldots ={\bf o}_8=0$}
\revised{
and $c_{1,2}$ corresponding to $c_1$ with ${\bf o}_1=\ldots ={\bf o}_4=0$.}

\revised{This results in an $EC(12,8)$, whose codewords look like
\[
[{\bf o}_1,\ldots,{\bf o}_8,c_{1,1},c_{1,2},c_2,c_3]
\]
where $c_{1,1}+c_{1,2}$ is equal to the original code's $c_1$:}

\revised{\[
c_{1,1}+c_{1,2} =c_1.
\]}

\colortext{For both Hierarchical and Pyramid codes, at each hierarchy level, there is some `local redundancy' which can repair lost blocks without accessing blocks outside the subgroup, while if there are too many errors within a subgroup, then the `global redundancy' at that level will be used. One moves further up the pyramid until repair is eventually completed. Use of local redundancy means that a small number of nodes is contacted, which translates into a smaller bandwidth footprint. Furthermore, if multiple isolated (in the hierarchy) failures occur, they can be repaired independently and in parallel.}

\revised{In contrast to Hierarchical codes, where analysis of the resilience has not been carried out, Pyramid codes' top-down approach allows to discern distinct failure regimes under which data is recoverable, and regimes when data is not recoverable. For instance, in the example above, as long as there are three or fewer failures, the object is always reconstructable. Likewise, if there are five or more failures, then the data cannot be reconstructed. However, there is also an intermediate region, in this simple case, it being the scenario of four arbitrary failures, in which, for certain combinations of failures, data cannot be reconstructed, while for others, it can be.}

Coinciding with these works, researchers from the network coding
community started studying the fundamental limits and
trade-offs of bandwidth usage for regeneration of \bluetext{a} lost encoded
block vis-a-vis the storage overhead (subject to the MDS constraint) culminating in a new family of codes, broadly
known as \emph{regenerating codes}, discussed next.

\section{Regenerating codes}

The repair of lost redundancy in a storage system can be abstracted
as an \emph{information flow graph} \cite{DGWK-journal}.

\begin{figure}[htbp]
    \begin{center}
    \subfigure[\label{fig:rgccloud}
    \colortext{
    Information flow graph for regenerating codes: each storage node is modeled as two virtual nodes, $X_{in}$ which collects $\beta$ amount of information from arbitrary $d$ live nodes, while storing a maximum of $\alpha$ amount of information, and  $X_{out}$, which is accessed by any data collector contacting the storage node.} A max-flow min-cut analysis yields the feasible values for storage capacity $\alpha$ and repair bandwidth $\gamma=d\beta$ in terms of the number $d$ of nodes contacted and code parameters $n,k$, where $d \geq k$.]{\includegraphics[scale=0.5]{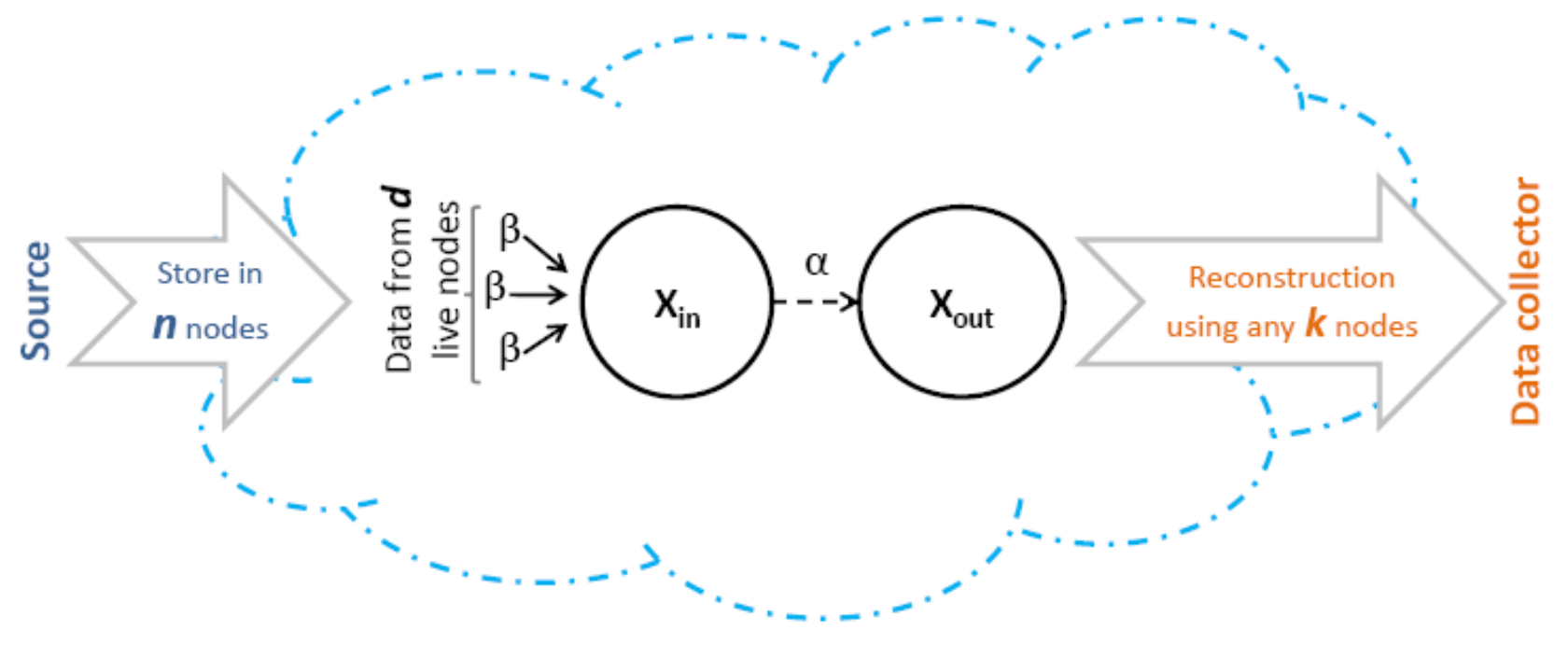}}
    \subfigure[\label{fig:RGCtradeoffcurve}Trade-off curve for the amount of storage space $\alpha$ used per node, and the amount of bandwidth $\gamma$ needed to regenerate a lost node. If multiple repairs $t$ are carried out simultaneously, and the $t$ new nodes at which lost redundancy is being created collaborate among themselves, then better trade-offs can be realized, as can be observed from the plot \colortext{(done using $k=32$, $d=48$, $n$ can be any integer bigger than $d+t$).}]{\includegraphics[scale=0.5]{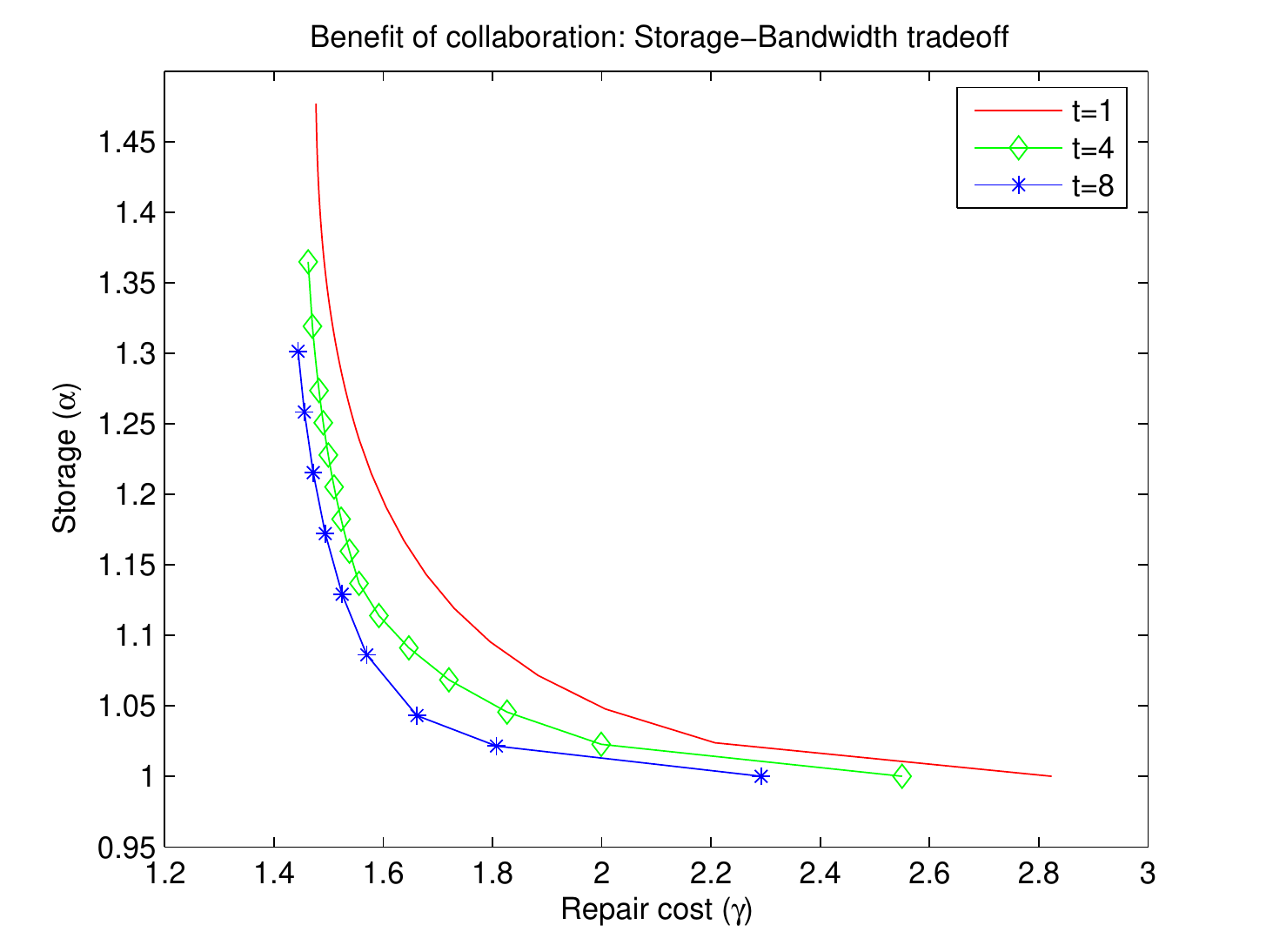}}
  \end{center}
  \caption{The underlying network coding theory inspiring regenerating codes.}
  \label{fig:RGCcloudandtradeoff}
\end{figure}
\colortext{
First, the data object is encoded using an $EC(n,k)$ MDS code and the encoded
blocks are stored across $n$ storage nodes. Each storage node is assumed
to store an amount $\alpha$ of data (meaning that the size of an encoded
block is at most $\alpha$, since only one object is stored).}
When one node fails, new nodes contact $d\geq k$ live nodes and download $\beta$ amount
of data from each contacted node in order to perform the repair.
\colortext{
If several failures occur, the model \cite{DGWK-journal} assumes that repairs are taken care of one at a time.
Information flows from the data owner to the data collector as follows (see Figure \ref{fig:rgccloud} for an illustration):}
(1) The original placement of the data distributed over $n$ nodes is modeled as directed edges of weight $\alpha$ from the sources (data owners) to the original storage nodes.
\colortext{
(2) A storage node is denoted by $X$ and modeled as two logical nodes $X_{in}$ and $X_{out}$, which are connected with a directed edge $X_{in}\rightarrow X_{out}$ with weight $\alpha$ representing the storage capacity of the node.
The data flows from the data owner to $X_{in}$, then from $X_{in}$ to $X_{out}$. (3) The regeneration process consists of directed edges of weight $\beta$ from $d$ contacted live nodes to the $X_{in}$ of the newcomer.}
(4) Finally, the reconstruction/access \colortext{of the whole object} is abstracted with edges of weight $\alpha$ to represent the destination (data collector) downloading data from arbitrary $k$ live storage nodes. Then, the maximum information that can flow from the source to the destination is determined by the max-flow over a min-cut of this graph. For the original \colortext{object} to be reconstructible at the destination, this flow needs to be at least as large as the size of the original object.

\begin{figure}[htbp]
    \begin{center}
    \subfigure[\label{fig:funcRGC}An example of functional repair for $k=2$ and $n=4$, adapted from \cite{DGWK-journal}: an object is cut into 4 pieces ${\bf o}_1,\ldots,{\bf o}_4$, and two linear combinations of them are stored at each node. When the 4th node fails, a new node downloads linear combinations of the two pieces at each node (the number on each edge describes what is the factor that multiplies the encoded fragment), from which it computes two new pieces of data, different from those lost, but any $k=2$ of the 4 nodes permit object retrieval. ]{\includegraphics[scale=0.35]{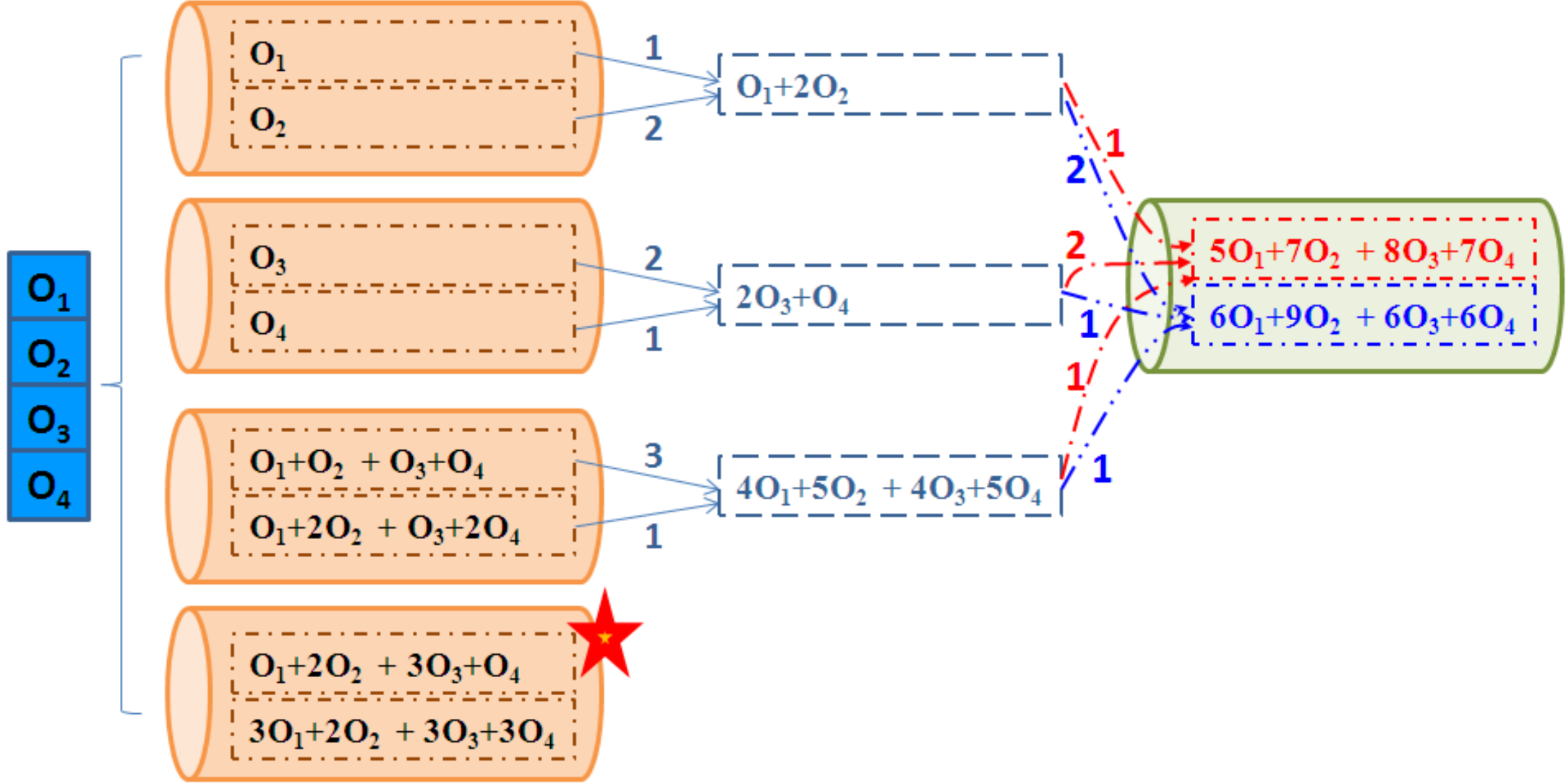}} \hspace{2mm}
    \subfigure[\label{fig:exactRGC}An example of exact repair from \cite{RSVR-allerton09}: an object ${\bf o}$ is encoded by taking its inner product with 10 vectors ${\bf v}_1,\ldots,{\bf v}_{10}$, to obtain ${\bf o}^T{\bf v}_i$, $i=1,\ldots,10$, as encoded fragments. They are distributed to the 5 nodes $N_1,\ldots,N_5$ as shown. Say, node $N_2$ fails. A newcomer can regenerate by contacting every node left, and download one encoded piece from each of them, namely ${\bf o}^T{\bf v}_1$ from $N_1$, ${\bf o}^T{\bf v}_5$ from $N_3$, ${\bf o}^T{\bf v}_6$ from $N_4$ and ${\bf o}^T{\bf v}_7$ from $N_5$.]{\includegraphics[scale=0.56]{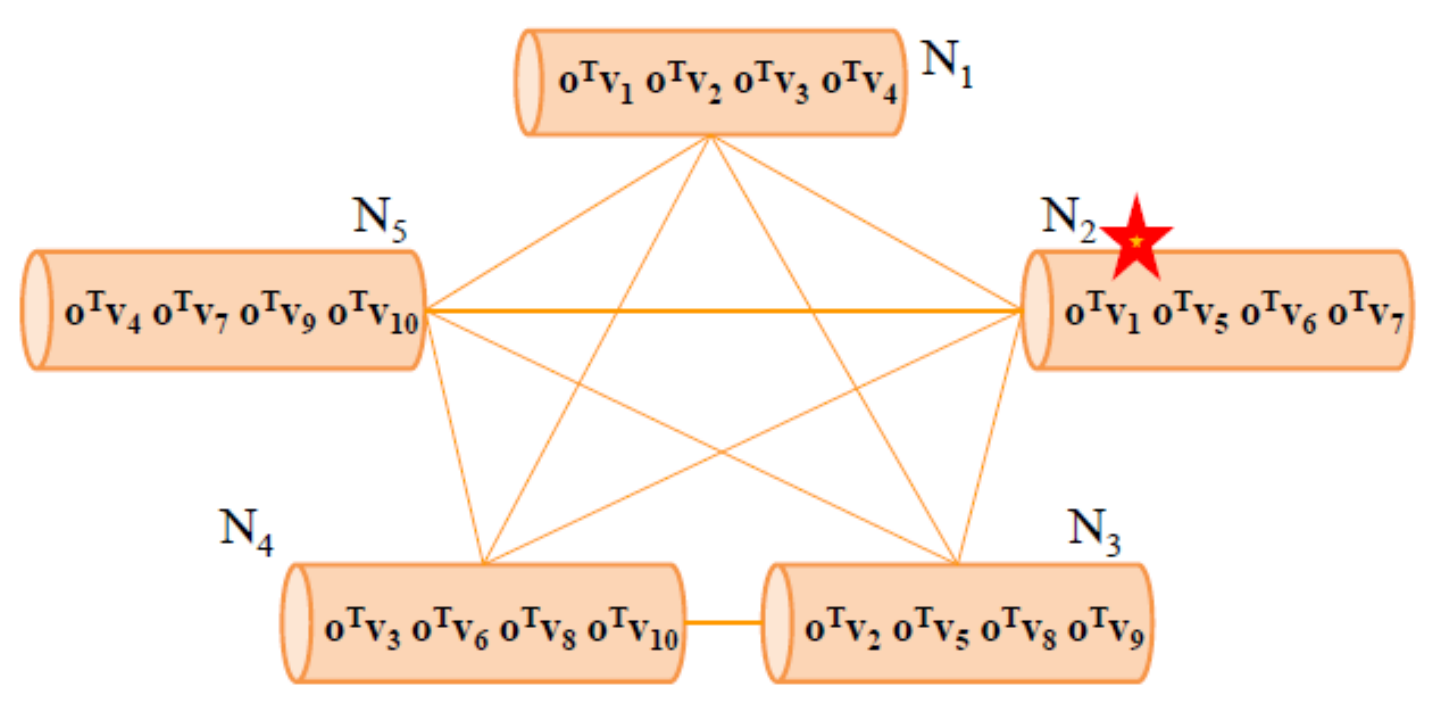}}
  \end{center}
  \caption{Regenerating codes: functional versus exact repair.}
  \label{fig:exampleRGCs}
\end{figure}
Any code that enables the information flow to be actually equal to the object size is called a regenerating code (RGC) \cite{DGWK-journal}. \colortext{Now, given $k$ and $n$, the
natural question is, \emph{what are the minimal
storage capacity $\alpha$ and bandwidth \bluetext{$\gamma=d\beta$} needed for repairing
an object of a given size}?} This can be formulated as a linear non-convex optimization
problem: minimize the total download bandwidth $d\beta$, subject to
the constraint that the information flow equals the object size. The
optimal solution is a piecewise linear function, which describes a
trade-off between the storage capacity $\alpha$ and the bandwidth
$\beta$ as shown in Figure \ref{fig:RGCtradeoffcurve} \bluetext{[$t=1$]}, and has two
distinguished boundary points: the minimal storage repair (MSR)
point (when $\alpha$ is equal to the object size divided by $k$),
and the minimal bandwidth repair (MBR) point.

\colortext{The trade-off analysis only determines what can best be achieved, but in itself
does not provide any specific code construction. Several codes
have since been proposed, most of which operate either at the MSR or MBR points of the trade-off curve, e.g., \cite{RSVR-allerton09}.}

\colortext{The specific codes need to satisfy the constraints determined by the max-flow min-cut analysis, however there is no constraint or need to regenerate precisely the same (bitwise) data as was lost (see Figure~\ref{fig:exampleRGCs} (a)). When the regenerated data is in fact not the same as that lost, but nevertheless provides equivalent redundancy, it is called \emph{functional regeneration}, while if it is bitwise identical to what was lost, then it is called \emph{exact regeneration}, as illustrated in Figure~\ref{fig:exampleRGCs} (b). \colortext{Note that the proof of storage-bandwidth trade-off determined by the min-cut bound does not depend on the type of repair (functional/exact).}}

The original model \cite{DGWK-journal} has since been generalized \cite{KLS,Shum-ICC} to show that in case of multiple faults, the new nodes carrying out regenerations can collaborate among themselves \colortext{to perform several repairs in parallel, which was in turn shown} to reduce the overall bandwidth needed per regeneration (Figure \ref{fig:RGCtradeoffcurve} \colortext{\bluetext{$t>1$ representing the number of failures/new collaborating nodes}}).
Instances of codes for this setting, \colortext{referred to as collaborative regenerating codes (CRGC)} are rarer than classical regenerating codes, and up to now, only a few code constructions are known \cite{Shum-ICC,Shum}.

\colortext{(Collaborative) regenerating codes stem from a precise information theoretical characterization. However, they also suffer from algorithmic and system design complexity inherited from network coding, which is larger than even traditional erasure codes, apart from the added computational overheads. The value of fan-in $d$ for regeneration has also practical implications. With a high fan-in $d$ even a small number of slow or overloaded nodes can thwart the repairs.}


\section{\revised{Locally repairable} codes}

\revised{The codes proposed in the context of network coding aim at reducing the repair bandwidth, and can be seen as the combination of an MDS code and a network code.
Hierarchical and Pyramid codes instead tried to reduce the repair degree or fan-in (i.e., the number of nodes needed to be contacted to repair) by using ``erasure codes on top of erasure codes". We next present some recent families of locally repairable codes (LRC) \cite{OD-infocom,src,OD-itw,ankit}, which minimize the repair fan-in $d$, trying to achieve $d<<k$ such as $d=2$ or $3$. Forcing the repair degree to be small has advantages in terms of repair time and bandwidth, however, it might affect other code parameters (such as its rate, or storage overhead). We will next elaborate a few specific instances of locally repairable codes.}

\revised{The term ``locally repairable'' is inspired by \cite{gopalan}, where the repair degree $d$ of a node is called the ``locality $d$" of a codeword coordinate, and is reminiscent of \emph{locally decodable} and \emph{locally correctable} codes, which are well established topics of study in theoretical computer science. Self-repairing codes (SRC) \cite{OD-infocom,src} were to our knowledge the first $(n,k)$ codes designed to achieve $d=2$ per repair for up to $\frac{n-1}{2}$ simultaneous failures. Other families of locally repairable codes based on projective geometric construction (Projective Self-repairing Codes) \cite{OD-itw} and puncturing of Reed-Mueller codes \cite{ankit} have been very recently proposed. Some instances of these latter codes can achieve a repair degree of either 2 or 3.}

\begin{figure}[htbp]
    \begin{center}
        \subfigure[\label{fig:hsrc}An example of self-repairing codes from \cite{OD-infocom}: 
    the object ${\bf o}$ has length 12, and encoding is done by taking linear combinations of the 12 pieces as shown, which are then stored at 7 nodes.]{\includegraphics[scale=0.32]{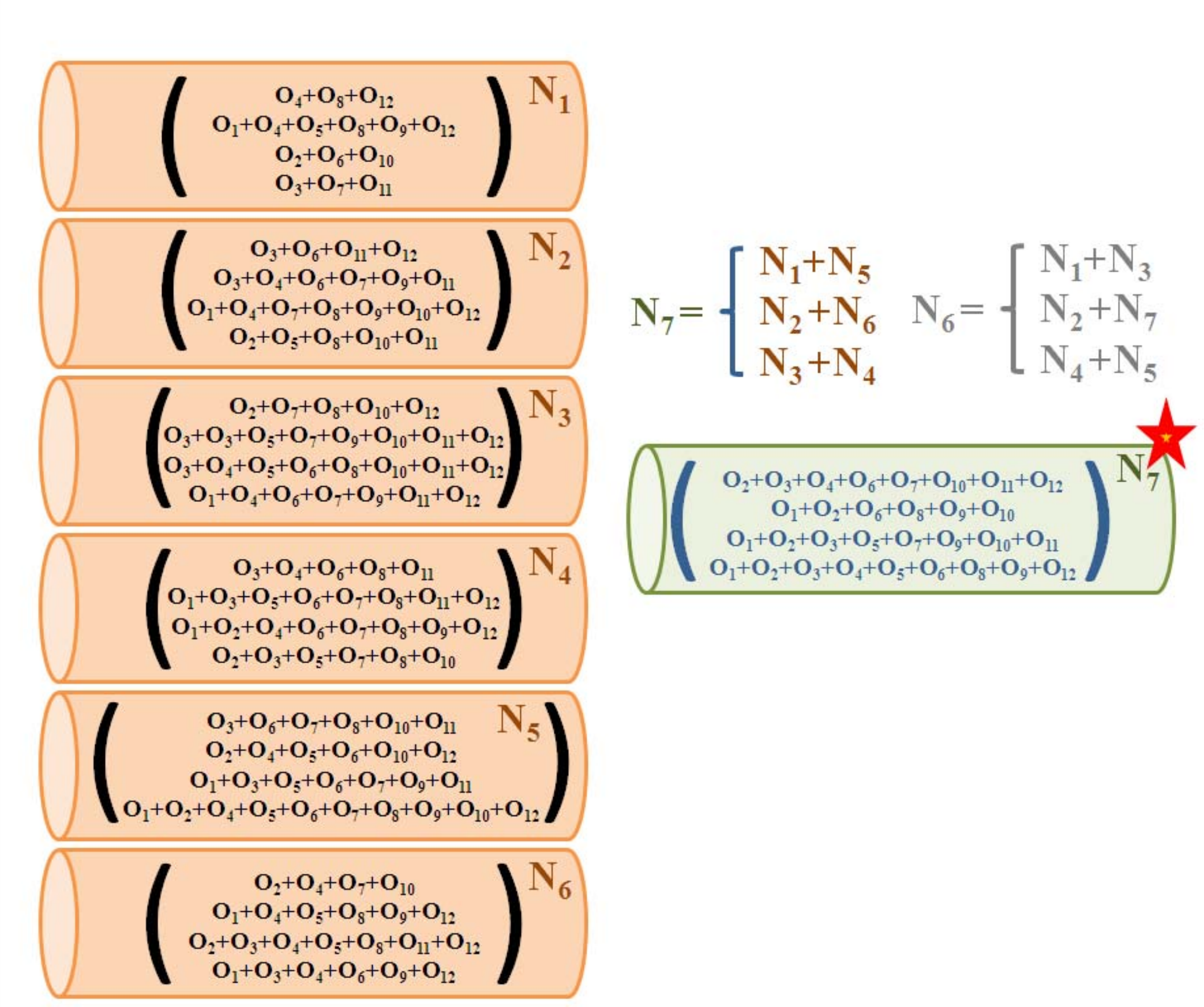}} \hspace{2mm}
    \subfigure[\label{fig:psrc}An example of self-repairing codes from \cite{OD-itw}: The object ${\bf o}$ is split into four pieces, and \emph{xor}-ed combinations of these pieces are generated. Two such pieces are stored at each node, over a group of five nodes, so that contacting any two nodes is adequate to reconstruct the original object. Furthermore, \emph{systematic} pieces are available in the system, which can be downloaded and just appended together to reconstruct the original data.]{\includegraphics[scale=0.4]{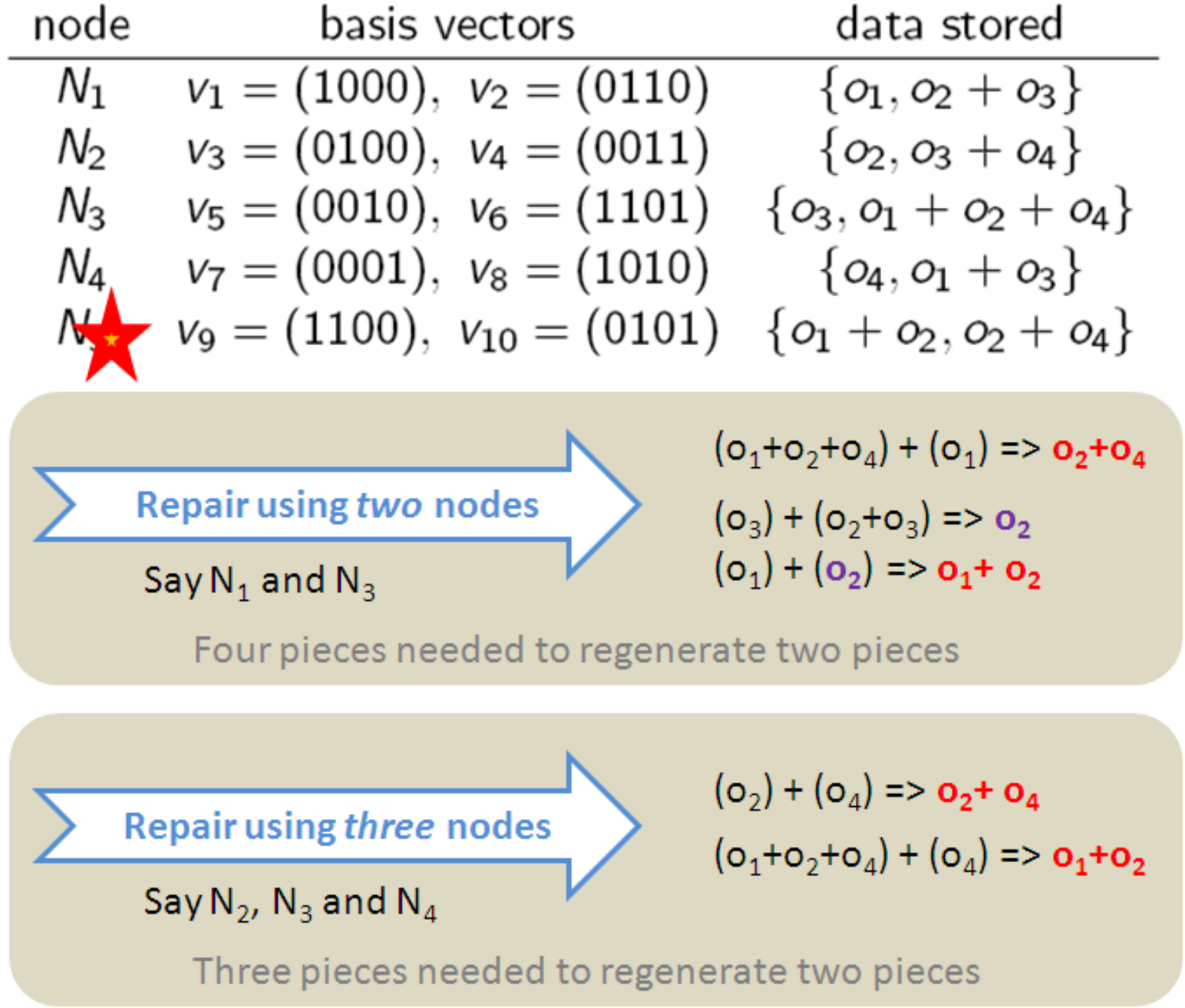}}
  \end{center}
  \caption{Self-repairing codes.}
  \label{fig:SRC}
\end{figure}

\colortext{With $d=2$ resources of at most two live nodes may get saturated due to a repair. Thus simultaneous repairs can be carried out in parallel, which in turn provides fast recovery from multiple faults. For example, in Figure~\ref{fig:hsrc} if the 7th node fails, it can be reconstructed in 3 different ways, by contacting either $N_1,N_5$, or $N_2,N_6$, or $N_3,N_4$. If both the 6th and 7th node fail each of them can still be reconstructed in two different ways. One newcomer can contact first $N_1$ and then $N_5$ to repair $N_7$, while another newcomer can in parallel contact first $N_3$ then $N_1$ to repair $N_6$.}

\colortext{Figure~\ref{fig:psrc} shows another example illustrating how the fan-in can be varied to achieve different repair bandwidths while using SRC. If a node, say $N_5$, fails, then the lost data can be reconstructed by contacting a subset of live nodes. Two different strategies with different fan-ins $d=2$ and $d=3$ and correspondingly different total bandwidth usage have been shown to demonstrate some of the flexibilities of the regeneration process.}

\colortext{Notice that the optimal storage-bandwidth trade-off of regenerating codes
does not apply here, since the constraint $d>k$ is relaxed. Thus
better trade-off points in terms of total bandwidth usage for a repair can also
be achieved (not illustrated here, see \cite{OD-itw} for details).}

\revised{Recall that if a node can be repaired with $d<k$ other nodes then there exist dependencies among them. The data object can be recovered only out of $k$ independent encoded pieces, and hence when the $k$ nodes include $d+1$ nodes with mutual dependency, then the data cannot be recovered from them. LRCs however allow recovery of the whole object using many specific combinations of $k$ encoded fragments. From the closed form and numerical analyses of \cite{OD-infocom} and \cite{OD-itw}, respectively, one can observe that while there is some deterioration of the static resilience\footnote{Static resilience is a metric to quantify a storage system's ability to tolerate failures based on its original configuration, and assuming that no repairs to compensate for failures are carried out.} with respect to MDS codes of equivalent storage overhead, the degradation is rather marginal. This can alternatively be interpreted as that for a specific desired value of fault-tolerance, the storage overhead for using LRC is negligibly higher than MDS codes. An immediate caveat emptor that is needed at this juncture is that, the rates of the known instances of locally repairable codes in general, and self-repairing codes in particular, are pretty low, and much higher rates are desirable for practical usage. The static resilience of such relatively higher rate locally repairable codes, if and when such codes are invented, will need to be revisited to determine their utility. Such trade-offs are yet to be fully understood, though some early works have recently been carried out \cite{gopalan,henk}.}

\section{\revised{Cross-Object Coding}}

\revised{All the coding techniques we have seen so far address the repairability problem at
the granularity of isolated objects that are stored using erasure
coding. However, a simple heuristic of superimposing two codes, one over individual objects, and another across encoded pieces
from multiple objects \cite{apsysDO} as shown in Figure
\ref{fig:RAID4onEC}, can provide good repairability properties as
well.}

\revised{Consider $m$ objects $O_1,\ldots,O_m$ to be stored. For $j=1,\ldots,m$, object
$O_j$ is erasure encoded into $n$ encoded pieces
$e_{j1},\ldots,e_{jn}$, to be stored in $m n$ distinct
storage nodes. Additionally, {\em parity groups} formed by $m$
encoded pieces (with one encoded piece chosen from each of the $m$
objects) can be created, together with a parity piece (or xor),
where w.l.o.g, a parity group is of the form $e_{1l},\ldots,e_{ml}$
for $l=1,\ldots,n$, and the parity piece $p_l$ is $p_l=e_{1l}+
\ldots + e_{ml}$. The parity pieces are then stored in additional $n$
distinct storage nodes. Such an additional redundancy is akin to
RAID-4.}

\revised{\begin{figure}[htbp]
    \begin{center}
  \includegraphics[scale=0.6]{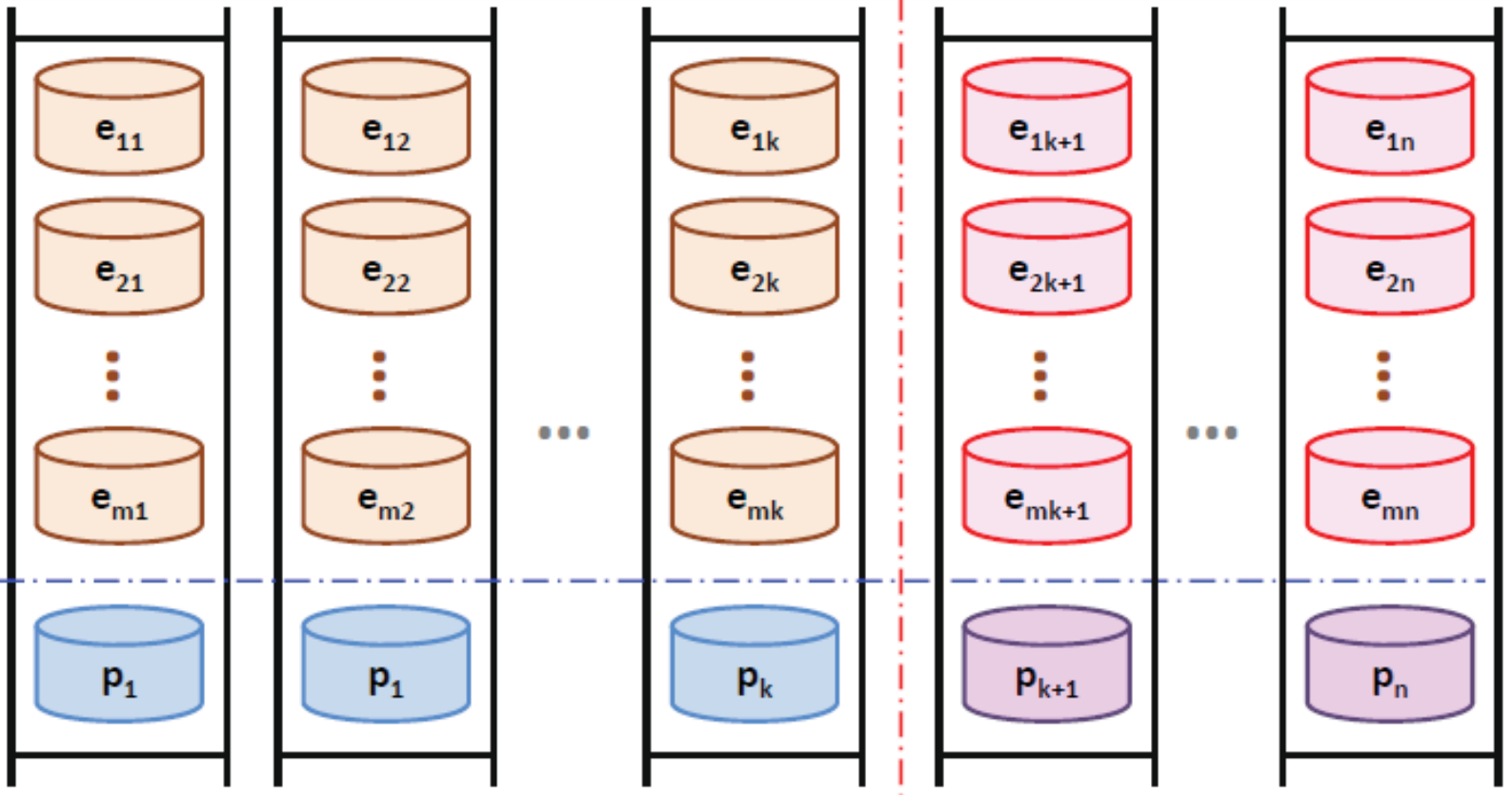}
  \end{center}
  \caption{Redundantly grouped coding: a horizontal layer of coding is performed on each object using an $(n,k)$ code, while a parity bit is computed vertically across $m$ objects, where $m$ is a design parameter.}
  \label{fig:RAID4onEC}
\end{figure}}

\revised{This code design, called \index{Redundantly grouped coding}{\em
Redundantly grouped coding} is similar to a two-dimensional product
code~\cite{Elias} in that the coding is done both horizontally and
vertically. In the context of RAID systems, similar strategy has also been applied to create intra-disk redundancy \cite{dholakia}. The design objectives here are somewhat different,
namely: (i) the horizontal layer of coding primarily achieves
fault-tolerance by using an $(n,k)$ erasure coding of individual
objects, while (ii) the vertical single parity check code mainly
enables cheap repairs (by choosing a suitable $m$) by creating
RAID-4 like parity of the erasure encoded pieces from different
objects.}

\revised{The number of objects $m$ that are cross-coded indeed determines the
fan-in for repairing isolated failures independently of the code
parameters $n$ and $k$. If $m<k$, it can be shown that the
probability that more than one failure occurs per column is small,
and thus repair using the parity bit is often enough - resulting in
cheaper repairs, while relatively infrequently repairs may have to
be performed using the $(n,k)$ code. The choice of $m$ determines
trade-offs between repairability, fault-tolerance and storage
overheads which have been formally analyzed in \cite{apsysDO}.
Somewhat surprisingly, the analysis demonstrates that for many
practical parameter choices, this cross-object coding achieves
better repairability while retaining equivalent fault-tolerance as
maximum distance separable erasure codes incurring equivalent
storage overhead.}

\revised{Such a strategy also leads to other practical concerns as well as
opportunities, such as the issues of object deletion or updates,
which need further rigorous investigation before considering them as
a practical option.}

\section{\revised{Preliminary comparison of the codes}}
\revised{The coding techniques presented in this paper have so far undergone only partial evaluation and benchmarking, and more rigorous evaluation of even the stand-alone approaches is ongoing work for most. Thus, it is somewhat premature to provide results from any comparative study, though some preliminary works on the same have also recently been carried out \cite{comparativestudy} taking into consideration realistic settings where multiple objects are collocated in a common pool of storage nodes, and multiple storage nodes may potentially fail simultaneously, all creating interferences between the different repair operations competing for the limited and shared network resources. Instead, we give one example of a theoretical result by considering the repair bandwidth per repair in the presence of multiple failures for some of these codes, and we provide an overview of what a system designer may expect from all these codes in Table-\ref{tab:parametersummary}. We further enumerate several other metrics that need to be studied to better understand their applicability.}

\begin{table*}
\begin{center}
\rowcolors{2}{gray!25}{white}
\begin{tabular}{p{3.5cm}p{3.5cm}cp{1.4cm}p{2cm}p{1.5cm}}
\rowcolor{gray!50}
code family& main design objective & MDS & fan-in $d$ & simultaneous repairs & bandwidth per repair \\
EC/RS & noisy channels & yes & $k$ & $\leq n-k$ & $1+\frac{k-1}{t}$ \\
RGC \cite{DGWK-journal}        & min. repair bandwidth & yes & $\geq k$ & 1 & $\frac{d}{d-k+1}$ \\
CRGC \cite{Shum-ICC,KLS}     & min. repair bandwidth & yes  & $\geq k$  & $t$ & $\frac{d+t-1}{d-k+1}$ \\
SRC \cite{OD-infocom}      & min. fan-in  & no & 2 & $\leq\frac{n-1}{2}$ & 2 \\
Pyramid \cite{pyramid}      & localize repairs (probabilistically) & no & depends & depends & depends \\
Hierarchical \cite{hierarchical} & localize repair (probabilistically) & no & depends & depends & depends \\
Cross-object coding \cite{apsysDO} & constant repair fan-in (probabilistically) & no & depends: $m$ or $k$ & depends & depends: $m$ or $k$ \\
\end{tabular}
\caption{\label{tab:parametersummary}Code design overview: \revised{We specify `depends' to some of the metrics, to signify that the corresponding value depends on the specific fault pattern and possible code parameters. A case in point being general Pyramid or Hierarchical codes. They have several parameters, the details of which we have not delved into in this high level survey. But, one can already note from the simple Hierarchical code example discussed in this paper that parallel repairs may be possible sometimes (for instance when ${\bf o}_1$ and ${\bf o}_4$ fail simultaneously), while it may have to be done in a serialized manner (for example, if ${\bf o}_1$ and ${\bf o}_1+{\bf o}_2$ fail simultaneously), while it may be impossible in other scenarios (such as when ${\bf o}_1$ and ${\bf o}_2$ fail simultaneously). The other aspects of repair likewise may vary, depending on failure pattern as well as code parameters.}}
\end{center}
\end{table*}

\begin{figure}[htbp]
    \begin{center}
\includegraphics[scale=0.5]{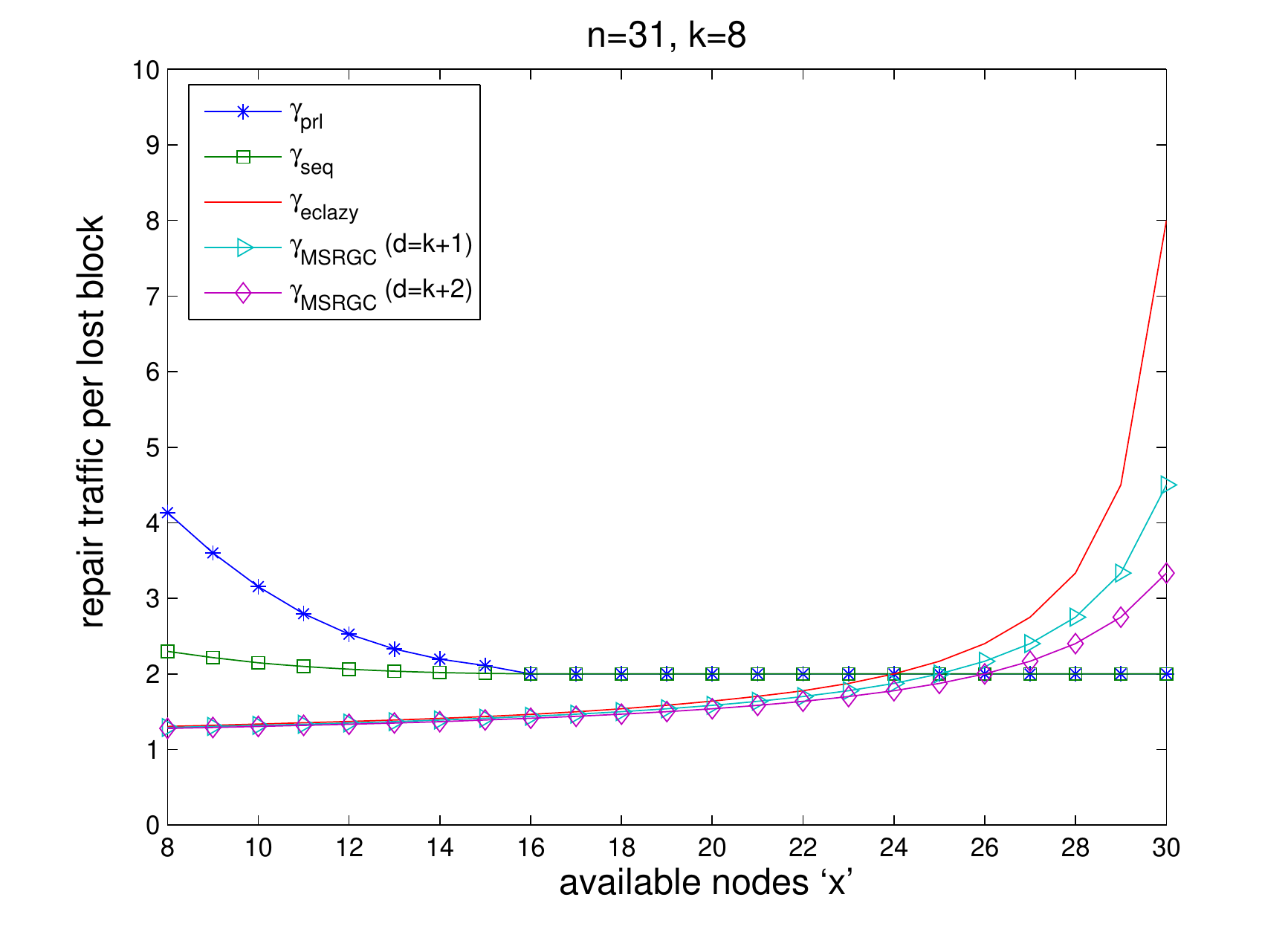}
  \end{center}
  \caption{\revised{Comparison among traditional erasure codes, regenerating codes and self-repairing codes (derived theoretically in \cite{OD-infocom}):} Average traffic normalized with $B/k$ per lost block for various choices of $x$ ($B$ is the size of the stored object) for \colortext{(n=31,k=8)} encoding schemes. For parallel repairs using erasure codes the traffic is $k=8$ (not shown). \colortext{The SRC code parameters are denoted as SRC(n,k).}}
  \label{fig:gamma}
\end{figure}

One would not allow in practice failures to accumulate indefinitely, and instead a regeneration process will have to be carried out. If this regeneration is triggered when precisely $x$ out of the $n$ storage nodes are still available, then the total bandwidth cost to regenerate each of the $n-x$ failed nodes is depicted in Figure \ref{fig:gamma}. \colortext{Note that delayed repair where multiple failures are accumulated may be a design choice, as in P2P systems with frequent temporary outages, or an inevitable effect of correlated failures where multiple faults accumulate before the system can respond.}

For locally repairable codes such as SRC the repairs can be done in sequence or in parallel, denoted $\gamma_{seq}$ and $\gamma_{prl}$ respectively in the figure. This is compared with MDS erasure codes ($\gamma_{eclazy}$) when the repairs are done in sequence, as well as with RGC codes at MSR point ($\gamma_{MSRGC}$) for a few choices of $d$. The bandwidth need has been normalized with the size of one encoded fragment. We notice that for up to a certain point, self-repairing codes have the least (and a constant of 2) bandwidth need for repairs even when they are carried out in parallel.

For larger number of faults, the absolute bandwidth usage for traditional erasure codes and regenerating codes is lower than that of self-repairing codes. However given that erasure codes and regenerating codes need to contact $k$ and $d\geq k$ nodes respectively, some preliminary empirical studies have shown the regeneration process for such codes to be slow \cite{PB} which can in turn make the system vulnerable. In contrast, because of an extremely small fan-in $d=2$, self-repairing codes can support fast and parallel repairs \cite{OD-infocom} while dealing with a much larger number of simultaneous faults. \revised{Comparison with some other codes such as hierarchical and pyramid codes has been excluded here due to the lack of necessary analytical results, as well as the fact that the different encoded pieces have assymetrical importance, and thus, just the number of failures does not adequately capture the system state for such codes.}

\revised{Given that repair processes run continuously or as and when deemed necessary, the static resilience is not the most relevant metric of interest for storage system designers. Often, another metric, namely \emph{mean time to data loss} (MTTDL) is used to characterize the reliability of a system. MTTDL is determined by taking into account the cumulative effect of the failures along with that of the repair processes. For the novel codes discussed in this manuscript, such study of MTTDL is yet to be carried out in the literature. However, a qualitative remark worth emphasizing is that, precisely because of the better repair characteristics such as fast repairs, some of these codes are likely to improve MTTDL significantly. Whether the gains outweigh the drawbacks, such as the lack of MDS property (and consequent poorer static resilience), is another open issue.}


\section{Concluding remarks}

There is a long tradition of using codes for storage
systems. This includes traditional erasure codes as well
as turbo and low density parity check codes (LDPC) coming from
communication theory, rateless (digital fountain and tornado) codes originally designed for content distribution
centric applications, or locally decodable codes emerging from the theoretical computer science community
to cite a few. The long believed mantra in applying codes for storage has been `\emph{the
storage device is the erasure channel}'.

Such a simplification ignores the maintenance process in NDSS for long term reliability. This realization has led to a renewed interest in designing codes tailor-made for NDSS. This article surveys the major families of
novel codes which emphasize primarily better repairability. \revised{There are many other system aspects which influence the overall performance of these codes, that are yet to be benchmarked. This high level survey is aimed at exposing the recent theoretical advances providing a single and easy point of entry to the topic. Those interested in further mathematical details depicting the construction of these codes may refer to a longer and a more rigorous survey \cite{ODlongsurvey} in addition to the respective individual papers. }

\section*{Acknowledgement}
A. Datta's work was supported by MoE Tier-1 Grant RG29/09. F. Oggier's work was supported by the Singapore National Research Foundation under Research Grant NRF-CRP2-2007-03.

\end{document}